\documentclass[%
 reprint,
 amsmath,amssymb,
 aps,
]{revtex4-2}

\usepackage{graphicx}
\usepackage{dcolumn}
\usepackage{bm}
\usepackage{xcolor}
\usepackage{tikz}
\usetikzlibrary{tikzmark, shapes, matrix, arrows.meta, bending, backgrounds, positioning, calc}
\usetikzlibrary{shapes.geometric, arrows.meta}
\usepackage[dvipsnames]{xcolor}
\usepackage{graphicx}
\usepackage{tcolorbox}
\usepackage{transparent}
\usepackage{booktabs} 
\usepackage{amsmath}  
\usepackage{amsmath}
\usepackage{array}      
\usepackage{ragged2e}   
\usepackage{booktabs}
\usepackage[table]{xcolor}
\usepackage{url}
\usepackage{hyperref}

\usepackage{xcolor}
\definecolor{C10}{HTML}{FF8C00} 
\definecolor{C11}{HTML}{191970} 
\definecolor{C13}{HTML}{DC143C} 
\definecolor{C12}{HTML}{9400D3} 

\definecolor{darkgreen}{rgb}{0.0, 0.5, 0.0}

\definecolor{delgray}{gray}{0.6}

\tikzset{
    flowstyle/.style={
        rectangle,
        draw,
        rounded corners,
        thick,
        text centered,
        minimum height=1.2cm,
        font=\sffamily
    },
    flowbox/.style={
        flowstyle,
        fill=white
    },
    flowfast/.style={
        flowstyle,
        fill=C13!80
    },
    flowprior/.style={
        flowstyle,
        fill=C11!75
    },
    flowresult/.style={
        flowstyle,
        fill=C12!70, 
    },
    flowslow/.style={
        flowstyle,
        fill=C10!80
    },
    flowarrow/.style={
        -Stealth, 
        thick
    }
}

\begin{document}

\preprint{APS/123-QED}

\title{Leveraging rapid parameter estimates for efficient gravitational-wave Bayesian inference via posterior repartitioning}

\author{Metha Prathaban}
\email{myp23@cam.ac.uk}
\affiliation{
 Kavli Institute for Cosmology, Madingley Road, Cambridge CB3 0HA, UK
}%
\affiliation{
Cavendish Laboratory, J.J. Thomson Avenue, Cambridge CB3 0HE, UK
}%

\author{Charlie Hoy}
\author{Michael J. Williams}%
\affiliation{%
 Institute of Cosmology and Gravitation, University of Portsmouth, Portsmouth, PO1 3FX, UK
}%

\date{\today}

\begin{abstract}
\vspace{0.2cm}

Gravitational-wave astronomy typically relies on rigorous, computationally expensive Bayesian analyses. Several methods have also been developed to perform rapid, approximate Bayesian inference. We present a novel approach to leverage the results of these low-latency analyses to accelerate the final inference, whilst ensuring that the Bayesian prior remains independent of the data. By combining the fast constraints from the \texttt{simple-pe} algorithm with the nested sampling acceleration technique of posterior repartitioning, we demonstrate that our method can guide the nested sampler towards the most probable regions of parameter space more efficiently for signal-to-noise ratios (SNR) greater than 20, while mathematically guaranteeing that the final inference is identical to that of a standard, uninformed analysis. We validate the method through an injection study on signals with SNR $<$ 150, demonstrating that it produces statistically robust and unbiased results whilst providing speedups of up to 210\%, with a mean speedup of 34\% for SNRs $>$ 20. Importantly, we show that the performance gain provided by our method scales with SNR, establishing it as a powerful technique to mitigate the cost of analysing signals from current and future gravitational-wave observatories.
\end{abstract}

\maketitle


\section{\label{sec:intro}Introduction}

A crucial part of gravitational-wave data analysis is parameter estimation (PE), the process of inferring physical properties of a source, such as its component masses or sky location, from the observed signal, given a model~\citep{LIGOScientific:2025yae}. The results of PE are fundamental to our understanding of stellar evolution, general relativity, and cosmology (most recently, see e.g.~\citep{LIGOScientific:2025pvj, LIGOScientific:2025jau, LIGOScientific:2021sio, Ray:2025xti,Hussain:2025llf,Tong:2025xir,Krishnendu:2025flx,Adamcewicz:2025phm,Stegmann:2025clo}). As the global network of gravitational-wave (GW) observatories improves in sensitivity and detects an ever-increasing number of events, the demand for efficient analysis methods is growing~\citep{aLIGO, aVirgo, aLVK_prospects, LIGOScientific:2014pky,VIRGO:2014yos,KAGRA:2020tym}, and will become more pressing as more GW detectors come online~\cite{dcc:M1100296}. This has led to several recent efforts in this area~\citep{Lange:2018pyp,Roulet:2024hwz,Pankow:2015cra,Pathak:2022iar,Tiwari:2023mzf,Nitz:2024nhj,Sharma:2025uva,Mushkin:2025yks,wangfstat}, including using machine learning algorithms~\citep{Gabbard:2019rde,Green:2020dnx,Green:2020hst,Dax:2021tsq,Dax:2024mcn,Raymond:2024xzj}. This new era of high signal-to-noise ratio (SNR) signals is exemplified by the recent loud event GW250114\_082203, which, with an estimated network SNR of $\sim 80$, is significantly louder than previous detections and promises a new frontier for precision science~\cite{LIGOScientific:2025rid,LIGOScientific:2025obp}. 

The standard approach to perform GW parameter estimation is through Bayesian inference~\citep{Thrane_2019, lal}. This is typically accomplished using stochastic sampling algorithms like nested sampling (NS)~\citep{Skilling:2004pqw, Skilling:2006gxv} (although see e.g. Refs.~\cite{Lange:2018pyp,Tiwari:2023mzf} for alternative techniques). Nested sampling is highly effective and reliable, providing not only estimates for the posterior distribution but also the Bayesian evidence required for model comparison~\citep{skilling}. However, the algorithm's computational cost presents a significant bottleneck. A standard PE analysis for a single event can require millions of likelihood evaluations, often translating to hours or days of runtime on high-performance computing clusters~\citep{LIGO_guide_signalextraction}.
This computational demand poses an impending challenge for the higher event rates of future observing runs, and already makes the routine use of the most physically comprehensive models, such as those including orbital eccentricity, expensive. 

To address the growing computational challenge associated with nested sampling, we introduce a method that combines two existing techniques: posterior repartitioning (PR)~\citep{PR1, PR2} and the \texttt{simple-pe} analysis~\citep{SimplePe1}. PR is a technique that can accelerate NS by using an approximate posterior, learned from a preliminary NS analysis, to define a more efficient sampling prior~\citep{Prathaban}. \texttt{simple-pe} leverages intuitive physical arguments to produce fast, probabilistic constraints on the source parameters~\citep{SimplePe1}, including general relativistic effects such as spin-induced orbital precession~\cite{Apostolatos:1994mx} and higher order multipole moments~\cite{Goldberg:1966uu,Thorne:1980ru}. Designed to produce initial parameter estimates in minutes on a single CPU, \texttt{simple-pe} can be employed for real-time alerts, as well as for informing more detailed parameter estimation offline. 

In this work, we propose a new framework that replaces the preliminary NS run previously used for PR with the rapid, physically-motivated constraints from \texttt{simple-pe}. This integration has the potential to be faster and more robust than previous implementations~\citep{Prathaban}, as it avoids a potentially slow and unreliable preliminary stochastic search. The result is a parameter estimation framework that significantly reduces the number of likelihood evaluations required for convergence, enabling more efficient analysis of both current and future data. Combining our method with techniques that accelerate the single likelihood evaluation time would further improve the performance gains~\citep{Cornish:2021lje, Canizares:2014fya, Multibanding1, Multibanding2}.

This paper presents and validates this new PE framework. First, we conduct a large-scale injection study on a population of signals representative of current detections to validate the statistical integrity of the method. We show that it produces unbiased posteriors that are consistent with those from standard NS analyses. Second, we perform a study to characterise how the method's performance scales with network SNR. This establishes the efficiency gains that can be expected, particularly for the loud events that will become more common in future observing runs.

The remainder of this paper is organised as follows. After a review of Bayesian inference and nested sampling in Section~\ref{sec:background}, we detail our PE framework in Section~\ref{sec:methods}. We then present two distinct studies to assess the method. In Section~\ref{sec:injection_study}, we validate its statistical integrity and in Section~\ref{sec:snr_scaling_study}, we characterise its performance scaling with SNR. Both sections detail the study design and present the corresponding results. Finally, we present conclusions and discuss future work in Section~\ref{sec:conclusions}.

\section{Background}
\label{sec:background}
\subsection{Bayesian Parameter Estimation in Gravitational Wave Astronomy}
\label{sec:background_bayes}

The Bayesian analysis of GW signals involves determining the properties of the astrophysical source, described by a set of parameters $\boldsymbol{\theta}$, given the observed strain data $\mathbf{d}$. Within the Bayesian inference framework, this is achieved by computing the posterior probability distribution, $\mathcal{P}(\theta|d, \mathfrak{M})$, which represents probability of the parameters given the data and a particular underlying model, $\mathfrak{M}$~\citep{Thrane_2019}. According to Bayes' theorem, the posterior is given by:
\begin{equation}
\label{eq:bayes_theorem}
\mathcal{P}(\theta|d, \mathfrak{M}) = \frac{\pi(\theta|\mathfrak{M}) \mathcal{L}(d|\theta, \mathfrak{M})}{\mathcal{Z}(d|\mathfrak{M})}.
\end{equation}
The model remains unchanged throughout this study, so we will omit the explicit dependence on the model but we refer interested readers to e.g.~\cite{Ashton:2021cub,Hoy:2022tst,Hoy:2024vpc} for multi-model inference. Here, $\pi(\theta)$ is the prior probability distribution, encoding our pre-existing knowledge about the parameters. The term $\mathcal{L}(d|\theta)$ is the likelihood, which quantifies the probability of observing the data $d$ given a specific set of parameters. The denominator, $\mathcal{Z}(d)$, is the Bayesian evidence, obtained by integrating the product of the likelihood and prior over the entire parameter space:
\begin{equation}
\label{eq:evidence}
\mathcal{Z}(d) = \int \pi(\theta) \mathcal{L}(d|\theta) \, d\theta.
\end{equation}
While essential for comparing competing models within Bayesian methodology~\citep{skilling} (although this does not account model accuracy, see Refs.~\cite{Hoy:2022tst,Hoy:2024vpc}), the evidence is computationally expensive to calculate.


\subsection{Nested Sampling}

\begin{figure}
    \centering
    \def\svgwidth{0.4\textwidth}
    \hspace{-0em}
    \vspace{-14em}
    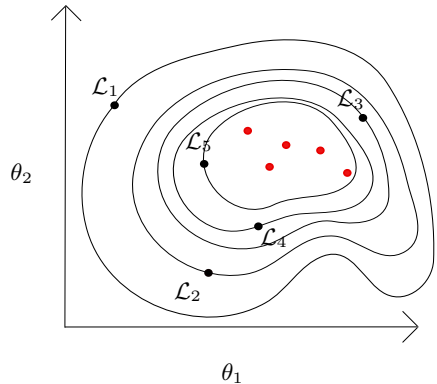
    \caption{Cartoon of nested sampling algorithm in two dimensions. Dead points define nested likelihood contours, indicated by $\mathcal{L}_{1}, \mathcal{L}_{2}, ...,$ in the parameter space, and the live points compress exponentially towards the peak of the likelihood as the algorithm proceeds. The fractional prior volumes enclosed between successive contours determine the weights of each dead point, used for calculating the Bayesian evidence and drawing posterior samples.}
    \label{fig:NS_cartoon}
\end{figure}

There exist several stochastic sampling methods to estimate the posterior distributions of GW data, such as MCMC~\citep{hastings1970, bilbymcmc} and nested sampling~\citep{skilling,lal, bilby_paper, dynesty, Williams2021Nessai}. NS is a Monte Carlo algorithm specifically designed to calculate the Bayesian evidence. NS works by drawing a set of `live points' from the prior distribution. At each iteration, the point with the lowest likelihood is removed from the live set, being made a `dead point', and is replaced with a new point drawn from the prior, subject to the constraint that its likelihood must be higher than that of the point it is replacing. This process generates a sequence of live points that iteratively move from the broad prior volume into progressively smaller regions of higher likelihood (see Figure~\ref{fig:NS_cartoon}). The collection of dead points, when appropriately weighted, provides samples from the posterior, while the sum of their likelihood values multiplied by their enclosed prior volume fractions yields a robust estimate of the evidence.

Despite its effectiveness, NS is computationally intensive. A single analysis often requires hours to days of processing time on high-performance computing clusters (although it has been shown that nested sampling can be parallelised over many CPUs~\cite{Smith:2019ucc}, reducing the overall wall time, and techniques have been developed to optimise the GW likelihood~\cite{Cornish:2010kf,Canizares:2014fya,Smith:2016qas,Vinciguerra:2017ngf,Zackay:2018qdy,Cornish:2021lje,Morisaki:2021ngj,Pathak:2022iar,Morisaki:2023kuq,Sharma:2025uva}, reducing the overall processing time). This cost is an obstacle to achieving rapid scientific turnaround for the large event catalogs of current and future observatories, assuming nested sampling continues to be used~\citep{aLVK_prospects, HuAccelerationReview}. For example, it has been shown that a full nested sampling algorithm with the future Laser Interferometer Space Antenna (LISA) observatory can take more than $10^{5}$ CPU hours~\cite{Hoy:2023ndx}. 

\section{Methods}
\label{sec:methods}

\subsection{Posterior Repartitioning}
\label{sec:background_pr}

One method to reduce the computational cost of nested sampling is posterior repartitioning, which improves sampling efficiency by exploiting a fundamental property of Bayesian inference~\citep{PR1, PR2, supernest, Prathaban}. The posterior distribution and Bayesian evidence depend only on the product of the likelihood and the prior, $\mathcal{L}(\theta) \pi(\theta)$. We are therefore free to redefine these two components for computational purposes, so long as their product remains invariant. 

\begin{figure}[t!] 
    \centering
    \begin{tikzpicture}
        \def\figwidth{0.35\textwidth}
        
        \draw [gray, thick] (0,0) rectangle (\figwidth, 0.6*\figwidth);
        
        \draw [C13, thick] (0.65*\figwidth, 0.3*\figwidth) rectangle (0.9*\figwidth, 0.5*\figwidth);

        \coordinate (posterior_center) at (0.78*\figwidth, 0.4*\figwidth);
        
        \begin{scope}[shift={(posterior_center)}, rotate=-50]
            
            \filldraw [draw=C12, fill=C12!30] (0,0) ellipse (0.6cm and 0.3cm);
            
            \filldraw [draw=C12, fill=C12!65] (0,0) ellipse (0.45cm and 0.2cm);
            
            \filldraw [draw=C12, fill=C12] (0,0) ellipse (0.25cm and 0.1cm);
            
            \node[black] at (0,-0.5) {$\mathcal{P}$};
        \end{scope}

        \node[black] at (0.05*\figwidth, 0.05*\figwidth) {$\pi$};
        \node[black] at (0.78*\figwidth, 0.54*\figwidth) {$\pi'$};
        
    \end{tikzpicture}
    \caption{Most of the computational time of nested sampling is spent in locating the bulk of the posterior, $\mathcal{P}$, within the much larger prior volume, $\pi$. In posterior repartitioning, this region of interest is first identified by an approximate method. The nested sampling run is then initialised within this smaller volume, $\pi'$, using a repartitioned likelihood (Eq.~\ref{eq:repartitioned_likelihood}) to ensure the resulting posterior and evidence are identical to a standard analysis.}
    \label{fig:PR_cartoon}
\end{figure}
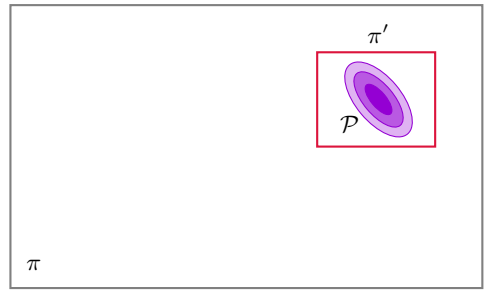

Nested sampling is uniquely suited to this manipulation because its algorithm explicitly separates the prior and the likelihood, by    `sampling from the prior subject to a hard likelihood constraint'. PR leverages this by redefining the prior that the NS algorithm sees. The procedure is as follows (also see Figure~\ref{fig:PR_cartoon}):
\begin{enumerate}
    \item An approximation to the target posterior, $g(\theta)$, is generated. In the past this has been achieved by training a normalizing flow on samples from a preliminary, low-resolution nested sampling run~\citep{Prathaban}. Normalizing flows learn a series of invertible mappings from a base distribution to a target~\citep{normalizingflow1, normalizingflows2}. 
    \item This learned distribution $g(\theta)$ is assigned as the new, effective prior for the NS algorithm. This is the \textbf{repartitioned prior}, $\pi'(\theta) \equiv g(\theta)$.
    \item To ensure the likelihood-prior product is unchanged, the likelihood must be redefined with a corrective term:
    \begin{equation}
    \label{eq:repartitioned_likelihood}
    \mathcal{L}'(\theta) = \mathcal{L}(\theta) \left[ \frac{\pi(\theta)}{\pi'(\theta)} \right]
    \end{equation}%
    The term in the brackets is the repartitioning factor. The nested sampler is then executed using $\pi'(\theta)$ as its sampling prior and $\mathcal{L}'(\theta)$ for its likelihood evaluations.
\end{enumerate}

It is important to note that this is purely a computational technique; the true Bayesian prior $\pi(\theta)$ of the inference problem is not altered. The result is an NS run that converges faster because its sampling is concentrated within the smaller volume of the repartitioned prior, which by design already resembles the posterior. The acceleration from PR fundamentally depends on the quality of the initial approximation, $\pi'(\theta)$. The approach of using a low-resolution NS run for initialization, as in~\cite{Prathaban}, presents two key drawbacks. First, the preliminary run itself can be computationally expensive, offsetting some of the efficiency gains. Second, and more critically, a stochastic sampler run in a low resolution configuration may fail to identify complex posterior structures, such as sharp peaks or secondary modes. If the initial approximation is poor, posterior repartitioning can be ineffective or, in the worst case, can cut off at the algorithmic level regions of the parameter space with non-negligible posterior probability, leading to biased results. Signals with lower signal-to-noise ratios (SNR) can be particularly susceptible to these issues.

We note that similar techniques that leverage preliminary or intermediate results have been developed for Markov Chain Monte Carlo and Sequential Monte Carlo \cite{Wolfe:2022nkv,SMC_accelerated}. We focus on PR, as nested sampling is the current standard for GW analyses. We leave a comparison to other approaches for further work. We also note that acceleration techniques which optimise the GW likelihood are complementary to PR and, whilst not considered, could also be applied.

\subsection{simple-pe}
\label{sec:background_simplepe}

The \texttt{simple-pe} algorithm provides a rapid analysis framework that generates estimates for the posterior probability distribution within minutes on a single CPU~\citep{SimplePe1}. Instead of performing a full Bayesian analysis with stochastic sampling, \texttt{simple-pe} uses simple, intuitive, and
physical insights to estimate the source properties. 

\texttt{simple-pe} takes advantage of the well-known physical decomposition of GW signals, whereby a GW signal can be decomposed into an infinite sum of spin-weighted spherical harmonics~\cite{Goldberg:1966uu,Thorne:1980ru}. However, \texttt{simple-pe} relies on the fact that only a small number of these terms significantly affect the observed signal; it can be shown that a GW signal consistent with general relativity (assuming quasi-circular orbits) can be described to high fidelity as a sum of three orthogonal components~\citep{SimplePe1,Hoy:2024ovd}. In essence, the dominant term describes the base chirp-like signal, the second term describes the contribution due to spin-induced precession and the third term describes the contribution from higher order multipole moments. The central principle of \texttt{simple-pe} is that a measurement of the dominant term provides the primary information for the intrinsic parameters: chirp mass, mass ratio, and aligned spins. A measurement of the precessing and/or higher order multipole terms then further constrains
these intrinsic measurements and provides information on the extrinsic properties. While \texttt{simple-pe} is capable of analysing precessing systems, with the possibility of incorporating eccentric binaries~\cite{Patterson:2024vbo}, this is not explored in this work. 

The \texttt{simple-pe} algorithm treats parameters hierarchically. It first identifies the most likely chirp mass, mass ratio and aligned-spin based on the best fitting dominant only-term that maximizes the matched filter SNR. Although a starting point is needed, which can come from a preliminary search, simple-pe has routines to determine a best-fitting template through an SNR optimisation step, similar to what is done in some online searches by the LVK~\citep{pycbc_live}. It then approximates the likelihood surface around this 3 dimensional peak as a multivariate Gaussian, where the width is based on
the expected accuracy at which the mass and spin can be measured, accounting for
known degeneracies throughout the parameter space~\cite{Owen:1995tm}. This simple distribution allows for the rapid generation of initial mass and spin samples. Extrinsic parameters, such as distance and orientation, are handled separately by sampling from constrained distributions that depend on the network SNR and response. Finally, SNRs for features beyond the dominant term are obtained by matched filtering the data again, and these initial samples are refined through a re-weighting step; each sample is assigned a weight based on the probability of obtaining each of the observed SNRs.

While this approach sacrifices the full rigour of an NS analysis, its physically-informed logic makes it a more robust tool for generating initial parameter estimates. Because its constraints are derived from specific physical features rather than stochastic exploration, it may be less susceptible to the same failure modes as a low-resolution sampler. For example, it is potentially less likely to miss secondary modes of the posterior, such as in the inclination angle, which is known to be often bimodal. This makes it an ideal candidate for improving the initialization stage of posterior repartitioning. 

However, due to the approximations employed, \texttt{simple-pe} in its current form may not guarantee the most reliable estimates. For example, for binaries with significant spin-precession or higher order multipoles, more than three terms are needed in the waveform expansion to fully describe the observed signal~\cite{Fairhurst:2019vut,McIsaac:2023ijd}. By neglecting some higher order multipole power, biases may be expected particularly in high mass systems (see e.g.~\cite{Ursell:2025ufb}). Similarly, by using only the dominant term to initially provide samples for the mass and spin of the binary, biases may be expected for high SNR signals with significant contributions from subdominant terms~\citep{Hoy:2024ovd} (although the initial samples are re-weighted to account for the observed power in subdominant terms; if the initial distribution lies outside of the region consistent with the observed power, the re-weighting procedure fails). Finally, \texttt{simple-pe} may struggle for low SNR signals since the multivariate Gaussian approximation for the initial mass and spin samples begins to break down.

\subsection{The \texttt{simple-pe}-PR Algorithm}
\label{sec:pipeline}

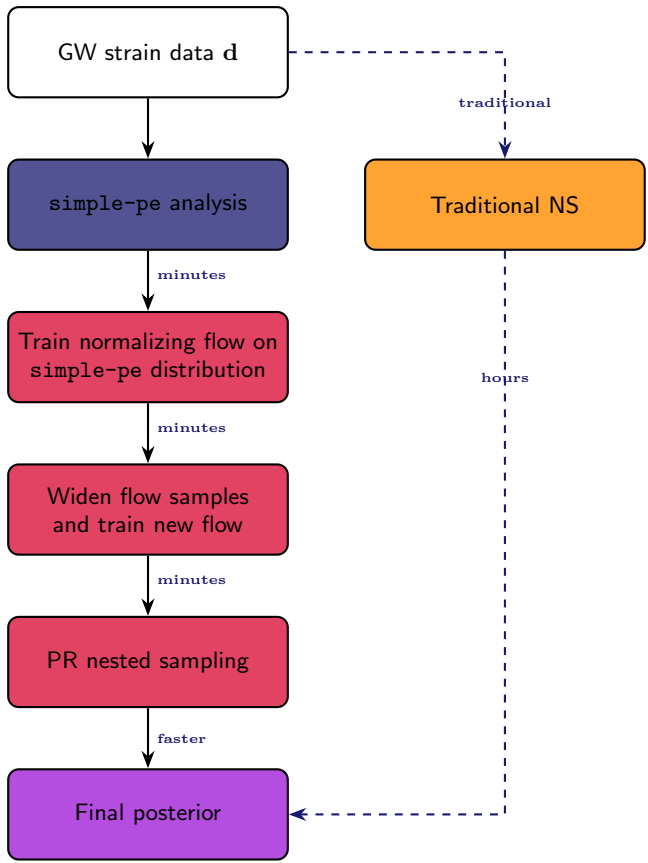
\begin{figure} [t!] 
    \centering
    \begin{tikzpicture}[node distance=0.8cm and 1cm]
        \node (data) [flowbox, text width=3.5cm] {GW strain data $\mathbf{d}$};
        
        \node (simplepe) [flowprior, below=of data, text width=3.5cm] {\texttt{simple-pe} analysis};
        
        \node (guess) [flowfast, below=of simplepe, text width=3.5cm] {Train normalizing flow on \texttt{simple-pe} distribution};

        \node (guess2) [flowfast, below=of guess, text width=3.5cm] {Widen flow samples and train new flow};
        
        \node (pr) [flowfast, below=of guess2, text width=3.5cm] {PR nested sampling};
        
        \node (final) [flowresult, below=of pr, text width=3.5cm] {Final posterior};
        
        \node (ns) [flowslow, right=of simplepe, text width=3.5cm] {Traditional NS};
        
        \draw [flowarrow] (data) -- (simplepe);
        \draw [flowarrow] (simplepe) -- node[right, font=\tiny, pos=0.4] {\textcolor{C11}{\textbf{minutes}}}(guess);
        \draw [flowarrow] (guess) -- node[right, font=\tiny, pos=0.4] {\textcolor{C11}{\textbf{minutes}}}(guess2);
        \draw [flowarrow] (guess2) -- node[right, font=\tiny, pos=0.4] {\textcolor{C11}{\textbf{minutes}}}(pr);
        \draw [flowarrow] (pr) -- node[right, font=\tiny] {\textcolor{C11}{\textbf{faster}}} (final);
        
        \draw [flowarrow, C11, dashed] (data.east) -- ++(1.75,0) -| node[above, font=\tiny, pos=0.8] {\textcolor{C11}{\textbf{traditional}}} (ns.north);
        \draw [flowarrow, C11, dashed] (ns.south) |- node[below, font=\tiny, pos=0.1] {\textcolor{C11}{\textbf{hours}}} (final.east);
        
    \end{tikzpicture}
    \caption{A flowchart of our proposed algorithm. Our method (left) introduces several fast pre-processing steps before running a posterior repartitioned (PR) nested sampling analysis. This reduces the runtime compared to a traditional nested sampling (NS) analysis (right). The primary computational bottleneck in both methods is the nested sampling step itself. Provided the initial \texttt{simple-pe} distribution is sufficiently widened, our method produces the same posteriors as the traditional approach but with significantly fewer likelihood evaluations.}
    \label{fig:method_flowchart}
\end{figure}

Our method integrates the constraints from \texttt{simple-pe} into the posterior repartitioning framework through a multi-step algorithm. The full procedure, shown schematically in Figure~\ref{fig:method_flowchart}, is as follows:

\begin{enumerate}
    \item \textbf{Initial Constraints:} The method begins by running \texttt{simple-pe} on the gravitational-wave strain data to generate a set of samples. These samples provide a rapid, physically-motivated initial approximation of the posterior.

    \item \textbf{Learning a Continuous Distribution:} The discrete samples from \texttt{simple-pe} must be converted into a continuous probability distribution from which we can both sample and evaluate the log-probability. We achieve this by training a normalizing flow, using a modified version of the \texttt{margarine} software~\cite{margarine1, margarine2, margarine_unbounded}, on the \texttt{simple-pe} samples (see Appendix~\ref{appendix:wideningflows} for details). This flow serves as a learned, continuous representation of the initial posterior approximation. However, other methods besides normalizing flows are possible for obtaining this, such as generation of a multi-dimensional kernel density estimate~\cite{2020SciPy-NMeth}, or Gaussian processes~\cite{DEmilio:2021laf}. 

    \item \textbf{Widening for Robustness:} Due to the approximations employed, the distribution produced by \texttt{simple-pe} can be narrower than, or offset from, the true posterior~\footnote{We note that \texttt{simple-pe} is still in active development and it is possible that in the future this widening procedure will not be required to obtain robust results.}. To prevent this from biasing the final inference, the learned distribution is widened. The specific widening factors applied are an important choice, and their empirical calibration based on our injection study is detailed in Section~\ref{sec:calib}.  The widening itself is achieved by modifying the base distribution of the trained normalizing flow. Specifically, we increase the standard deviation of the base multivariate Gaussian distribution. The covariance matrix is diagonal, meaning that we can change the standard deviation for each dimension independently. 2000 samples are then drawn from this `widened' flow. A second normalizing flow is then trained on these widened samples to produce the final, robust repartitioned prior, $\pi'(\theta)$. During training of both flows, we do not apply bounds and allow the flows to have support outside the original prior. This will not affect the final inference as samples outside the original prior will not be accepted by the sampler anyway, and we discuss this further in Appendix~\ref{appendix:wideningflows}. 

    \item \textbf{Final Parameter Estimation:} The final PE analysis is performed using \texttt{bilby-pr}, a custom \texttt{bilby} plugin package developed for this work and made publicly available on GitHub~\citep{bilby_pr}. This package extends the standard \texttt{dynesty} sampler~\citep{dynesty} as implemented in \texttt{bilby}~\citep{bilby_paper}, with arguments to accept a pre-trained \texttt{margarine} normalizing flow. It then implements custom prior and likelihood wrappers that employ the flow as the new repartitioned prior, $\pi'(\theta)$, and adjust the likelihood according to Eq.~\ref{eq:repartitioned_likelihood}.
\end{enumerate}

The repartitioned prior, $\pi'(\theta)$, does not incorporate information for all parameters. The constraints from \texttt{simple-pe} on the sky location parameters, right ascension ($\alpha$) and declination ($\delta$), exhibited a variance in accuracy that made them unsuitable for reliable repartitioning~\footnote{Although we note that \texttt{simple-pe} is still in active development and this will likely be improved in future versions.}. Similarly, \texttt{simple-pe} does not provide distributions for the phase, polarization angle, and time of coalescence. For these parameters, we retain the original priors. Finally, since we analytically marginalise over luminosity distance in the likelihood in order to improve convergence~\citep{Singer:2015ema, Singer:2016eax}, we do not use the \texttt{simple-pe} distance estimates. In summary, our repartitioned prior only incorporates information from \texttt{simple-pe} for the intrinsic binary parameters and the inclination angle, where its performance is most robust.

\section{Injections Study}
\label{sec:injection_study}

We conducted an injection study to validate the statistical robustness of \texttt{simple-pe}-PR on a population of simulated binary black hole merger signals, each 8 seconds in duration. 
The parameters for these injections were drawn from the prior distributions detailed in Table~\ref{tab:priors} in Appendix~\ref{appendix:priors}. This choice of priors is representative of those used in LVK analyses, including the choice of mass priors (see Appendix~\ref{appendix:priors} for further discussion). It therefore allows us to validate our method's statistical robustness in a similar parameter range.

The signals were generated using the aligned-spin waveform model \texttt{IMRPhenomXHM}, which includes higher order multipoles~\citep{PhenomXHM}. They were injected into coloured Gaussian noise representative of the design sensitivity of the Advanced LIGO-Hanford (H1), LIGO-Livingston (L1), and Virgo (V1) detectors~\cite{O4PSD}.

\subsection{Calibration of Widening Procedure}
\label{sec:calib}

 A key component of our method is the widening of the initial \texttt{simple-pe} distribution, by applying a multiplicative increase to the standard deviation of the flow's base distribution. This ensures the resulting posteriors are unbiased. The primary goal of this first study is to determine the requisite widening.



To perform an empirical calibration of the necessary widening factors, we drew a set of 100 injections from the prior. For each of these calibration signals, we compared the initial \texttt{simple-pe} distribution to the full posterior from a standard NS analysis. For each repartitioned parameter, we then determined the widening factor required to ensure the 1st and 99th percentiles of the reference one-dimensional marginal posterior were contained within the 1st and 99th percentiles of the one-dimensional marginal widened distribution. 

This calibration confirmed that the performance of \texttt{simple-pe} depends on the network SNR~\citep{SimplePe1}. A notable change in behavior was observed around an SNR of 20: \texttt{simple-pe} generally performed better on signals louder than this, and hence needed less widening. Consequently, we adopted two distinct sets of widening factors: one for signals in the lower-SNR regime (network SNR $<$ 20) and a less conservative set for high-SNR signals (network SNR $\ge$ 20), shown in Table~\ref{tab:widening_low_snr}.

The widening factors we use are intentionally conservative to ensure robustness for the majority of signals with SNRs above 10. For many signals, the initial distributions require little to no widening; 25\% of the injections in the calibration set needed widening factors less than 1.25 in all parameters. As discussed in Section~\ref{sec:snr_scaling_study}, this has an impact on the observed performance gains. For signals below SNR 10, the required widening becomes large enough to negate any computational benefit from posterior repartitioning. While we apply the lower-SNR factors to these signals for the purpose of our proceeding injection study, this fixed widening is often insufficient to guarantee an unbiased posterior and we do not recommend application of our method in this regime (see Section~\ref{sec:pp} for further discussion).


\begin{table}
\centering
\begin{tabular}{lc}
\toprule
\textbf{Parameter}     & \textbf{Widening Factor ($w$)} \\ \midrule
Mass Ratio ($q$)            & 2.5                   \\
Chirp Mass ($\mathcal{M}_c$)             & 2.0 (1.5)                 \\
Spin of lighter BH ($\chi_1$)         & 2.0 (1.5)                  \\
Spin of heavier BH ($\chi_2$)         & 1.0                   \\
Inclination angle ($\theta_\textrm{JN}$)           & 2.5                   \\ \bottomrule
\end{tabular}
\caption{Widening factors ($w$) applied to the standard deviation of the normalizing flow's base distribution for signals with network SNR $<$ 20 (SNR $\geq$ 20).}
\label{tab:widening_low_snr}
\end{table}

Finally, we note that these widening factors were calibrated on aligned-spin injections in a three-detector network. They may differ for analyses involving two-detector networks or precessing binary systems. We also expect these widening factors to vary with signal duration and prior choices (Appendix~\ref{appendix:priors}), but do not expect this to be the primary determinant in our method's performance (see Section~\ref{sec:snr_scaling_study}). The widening factors were observed to be largely independent of mass over the parameter range considered (where the primary mass spanned from 11.5 to 152.3$M_{\odot}$); other structures in the likelihood, such as those related to SNR, have a more significant impact. While this widening process could be automated in future work, using techniques similar to those introduced in~\citep{Prathaban}, we do not explore this here.

\subsection{Statistical Validation and Performance}
\label{sec:pp}

We now conduct a large-scale injection study to demonstrate that our method achieves correct statistical coverage and is consistent with the standard version of \texttt{bilby} and nested sampling~\citep{bilby_validate}. We also assess the method's performance. To perform the injection study, we drew a new set of 100 injections from the prior. For each of the signals, we performed two analyses: one using a standard nested sampling algorithm and one using the \texttt{simple-pe}-PR algorithm described in Section~\ref{sec:pipeline}. For the \texttt{simple-pe}-PR analyses, we applied the appropriate SNR-dependent widening factors from Table~\ref{tab:widening_low_snr}.

For both analyses, the nested sampling step was performed with the \texttt{dynesty} sampler~\citep{dynesty} as implemented in \texttt{bilby}. The right ascension ($\alpha$), polarization angle ($\psi$), and phase at coalescence ($\phi$) were treated with periodic boundary conditions. Identical sampler settings were used for both the standard and PR analyses to ensure a direct comparison; we used 1000 live points and the `\texttt{acceptance walk}' sampler, with its default settings~\citep{LIGOScientific:2025yae}. 


\subsubsection{Percentile-Percentile Comparison with Standard Nested Sampling}

The statistical integrity of our method is validated by a population-level comparison with the standard NS analysis, using the statistics of the probability-probability (P-P) plot~\citep{pplot}. A P-P plot shows the fraction of signals for which the injected parameter values lie within a given credible interval, as a function of the credible interval. In the ideal case where the parameter inference procedure leads to the correct coverage, this plot should be a diagonal line. However, there is a statistical uncertainty that arises from the finite number of signals analysed.

The recovered distributions for all parameters from our method are consistent with those from the normal NS method. We note that the individual P-P plots for both methods revealed small biases in the declination and time of coalescence at the geocentre parameters. Importantly, though, these biases do not arise from changes related to our method; our results are consistent with standard NS. We attribute these biases to statistical variance within this specific realisation of 100 injections, as they were not present in the separate realisation used for calibration. Sky location and time of coalescence at the geocentre have known degeneracies~\citep{Fairhurst_2009}, which may cause statistical fluctuations in one parameter to manifest as apparent biases in both. The P-P plots obtained from this set and the calibration set are included in Appendix~\ref{appendix:ppplot} for reference. It should also be noted that summary statistics, such as the P-P plot, have been previously shown to be subject to an additional source of uncertainty that is not typically quantified in P-P plot error bars~\citep{costless_correction}, which could be relevant to these apparent biases; we did not include this source of uncertainty in our analysis, in line with standard practice in the literature. 

\begin{figure}[h!]
    \centering
    \includegraphics{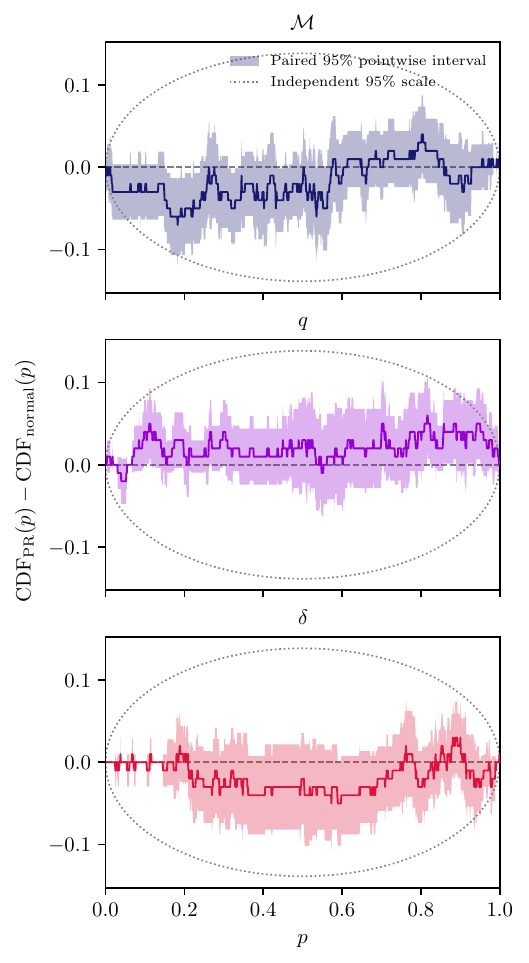}
    \caption{The solid lines show for each parameter the difference between the two P-P plots obtained from \texttt{simple-pe}-PR and standard NS, $\mathrm{CDF}_{\mathrm{PR}}(p) - \mathrm{CDF}_{\mathrm{normal}}(p)$, as a function of the credible interval, $p$. The ideal case for two methods which perfectly agree is a horizontal line at zero, but statistical uncertainties play a role. The dotted lines show the $2\sigma$ error bars if the two CDFs were statistically independent. This is shown for reference only, since in our case both methods run on the same set of injections so their P-P curves are positively correlated. The shaded bands therefore show the $2\sigma$ pointwise uncertainties estimated from the standard deviation of the differences. In particular, even the biased declination (bottom panel) shows consistency between the two methods, confirming that the biases are not due to changes related to \texttt{simple-pe}-PR.}
    \label{fig:pp_diff}
\end{figure}

To more directly compare the population level agreement between the methods, the differences between the two P-P plots for some representative parameters are shown in Figure~\ref{fig:pp_diff} (the remaining parameters are plotted in Figure~\ref{fig:pp_diff_remaining}). The $2\sigma$ error bars estimated from the injections are also plotted, which demonstrate again that the two methods are consistent with each other to within statistical fluctuations. 

\subsubsection{Event Level Agreement with Standard Nested Sampling}

For every injection, we also compute the individual one-dimensional Jensen-Shannon Divergence (JSD) between each pair of posterior distributions (see Appendix~\ref{appendix:jsd} for JSD methodology and full results, as well as~\cite{ourzenodo})~\citep{61115}. 
For the majority of injections, the posteriors from \texttt{simple-pe}-PR are in good agreement with those from standard NS, confirming the general robustness of the widening procedure. The JSD agreement also supports the hypothesis that the biases observed in the P-P plot are due to a statistical fluctuation in the properties of this specific set of injections. Nevertheless, we observed two injections above SNR 20 that exhibited significant differences in the posterior estimates. In both cases, this was due to \texttt{simple-pe} identifying the opposite sky mode, resulting in an offset inclination angle estimate. Our widening procedure is not designed to account for such cases where the initial estimate has significantly different support to the posterior. This type of failure occurs more frequently at lower signal strengths. Between SNR 10 and 20, six injections, representing 12.5\% of signals in this range, show significant posterior differences. Below SNR 10, this number increases to 10 of the 12 injections, an expected result as our widening factors were not calibrated for this regime.

Although the inclination angle failure mode can produce biased posteriors in rare instances, we still include it in our repartitioned prior to leverage the noticeable speedups it provides for the majority of signals. Further improvements to the \texttt{simple-pe} algorithm will alleviate this failure mode, both ensuring more robust results and reducing the need for our widening procedure. Moreover, the widening procedure itself could be improved by varying the degree of widening on the fly during the analysis, by sampling over an additional widening factor parameter, similarly to~\citep{Prathaban}. This would permit larger widening factors to be incorporated for more challenging signals without sacrificing performance on signals which require little to no widening. 

\subsubsection{Computational Performance}

\begin{figure}
    \centering
    \includegraphics{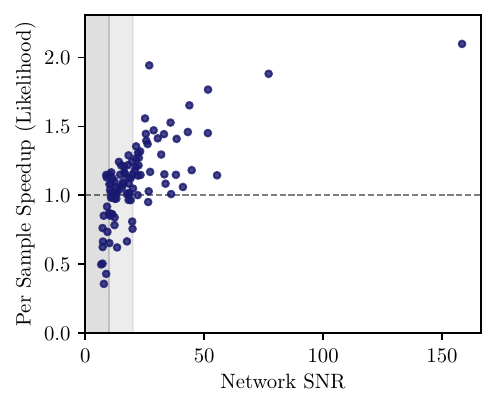}
    \caption{Per-sample likelihood evaluation speedup factor for each of the 100 injections, as a function of network SNR. Each point represents a single analysis. The lighter shaded region (SNR $<$ 20) highlights the regime where the speedup is unreliable. It also marks the boundary of the two sets of SNR-dependent widening factors. The darker shaded region (SNR $<$ 10) indicates where the calibrated widening factors may not be large enough to obtain unbiased results. The dashed line at 1.0 indicates no change in performance relative to a standard analysis.}
    \label{fig:injection_speedups_per_sample}

\end{figure}

We quantify the performance improvement, or speedup, of our method as the ratio of a given metric (e.g., number of likelihood evaluations) for a standard NS analysis to that of our \texttt{simple-pe}-PR analysis. The total runtime of our method is dominated by the nested sampling step. On average, \texttt{simple-pe} took $5$ CPU minutes, and training both the original and widened normalizing flows together took less than one CPU minute. By comparison, the nested sampling step took on average 7 hours, parallelised over 38 CPU cores. Therefore, when comparing runtimes, we only consider the respective nested sampling times of each method.
We could consider the speedup in terms of total likelihood evaluations, but one characteristic of PR is that the analysis typically spends a larger fraction of its runtime sampling the high-likelihood region of the parameter space, yielding a larger effective sample size than the equivalent standard analysis (see Appendix~\ref{appendix:ess}). Accordingly, our method can be performed with fewer live points than a standard NS analysis for the same final sample size. As such, rather than the total speedup, the per-sample speedup (the speedup factor normalized by the number of effective posterior samples from each analysis, calculated as in~\citep{Kish1995}) is the more informative performance metric, as it compares like-for-like inference products. 
Figure~\ref{fig:injection_speedups_per_sample} shows this per-sample speedup in terms of likelihood evaluations for each injection, as a function of network SNR. As anticipated from the expected performance of \texttt{simple-pe}~\cite{SimplePe1} and the SNR-dependence of the widening factors, there is a clear positive trend in performance with SNR, albeit with significant scatter. The maximum speedup observed is 2.1, for an injection with SNR 158. The performance distribution reflects the characteristics of our injection set, which was drawn from the prior distributions used in LVK analyses. Although the prior on the luminosity distance was narrowed to ensure that most signals have detectable SNRs (Table~\ref{tab:priors}), the majority of signals still have network SNRs below 20 (shaded region). The performance of our method in this regime is expected to be modest, or in some cases worse than standard NS. We find that 71\% of injections in our population show speedups greater than 1, and for signals with SNR $>$ 10, this figure increases to 78\%. The mean speedup for injections with SNR $>$ 20 was 1.34.  

As discussed in Section~\ref{sec:calib}, our chosen widening factors are intentionally conservative for the majority of signals above SNR 10; on the other hand, below SNR 10, the applied widening factors are generally expected to be insufficient. These results therefore highlight an important distinction between the statistical robustness and the computational efficiency of the method at different signal strengths.
For signals below SNR 10, the method is unlikely to produce reliable results and we do not recommend its use. For signals with SNR between 10 and 20, the empirically calibrated widening factors are robust for most cases, and the final inference will be statistically consistent with the standard analysis, excepting certain failure modes. However, as shown in Figure~\ref{fig:injection_speedups_per_sample}, the performance in this regime is unreliable, with some analyses resulting in a slowdown. Therefore, for both statistical validity and computational efficiency, we recommend the application of our method for the time being to signals with a network SNR $>$ 20. While this represents a minority ($\lesssim 10\%$) of current detections in GWTC-4.0, these high-SNR events are often the most information-rich signals and warrant repeated analyses to study carefully~\citep{GW190412, GW190521, GW231123}. As such, they account for a disproportionate fraction of the total computational budget for parameter estimation. For future observing runs, and observatories like the Einstein Telescope, signals with SNR $>$ 20 are predicted to constitute a larger fraction of the observed population; for Einstein Telescope approximately 50\% of all detections will have SNR $>$ 20~\citep{ET_SNR_pred}.

There is only one injection with SNR $>$ 20 that resulted in a marginal per-sample slowdown (5\% slower than standard NS); the final posterior distributions for this analysis were in excellent agreement with the standard NS analysis. The slowdown resulted from an over-broad repartitioned prior compared to the highly constrained target posterior, due to the conservative widening factors. Consequently, the internal random walk of the nested sampler frequently proposed points in regions of high repartitioned prior probability but low likelihood. This reduced the internal acceptance rate, thereby increasing the number of likelihood evaluations required to find a valid new live point at each nested sampling iteration. This event illustrates a key trade-off of our current design. The fixed widening procedure prioritises statistically unbiased inference, but is therefore occasionally too conservative, leading to localised sampling inefficiencies. Automating the widening procedure in future would lead to better performance improvements.

\subsubsection{Multimodality}

\begin{figure}
    \centering
    \includegraphics{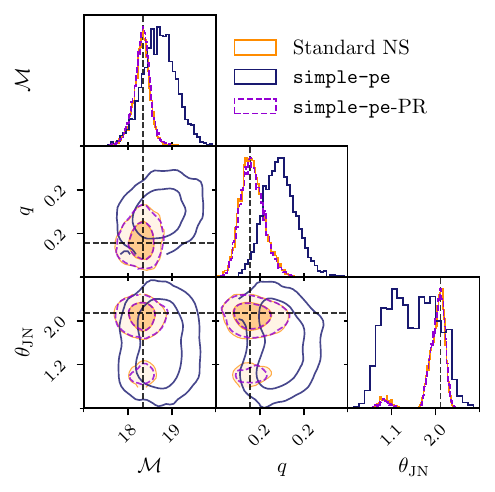}
    \caption{A representative injection from the study with a bimodal marginal posterior for the inclination angle ($\theta_\textrm{JN}$). The \texttt{simple-pe} approximation (unwidened) correctly identifies both modes of the posterior. This ensures the repartitioned prior encompasses the full posterior structure, including the secondary mode, leading to an unbiased final inference.}
    \label{fig:bimodal_example}
\end{figure}

A key advantage of \texttt{simple-pe} is that it is in general more robust to multi-modal posteriors, a common feature in gravitational-wave analysis that presents a significant challenge for other PR initialization methods. A low-resolution NS run or a Fisher matrix approximation can both fail to identify secondary posterior modes, particularly when they are well-separated from the primary mode. Such a failure leads to biased inference, as the repartitioned prior can inadvertently exclude regions of significant posterior probability. Figure~\ref{fig:bimodal_example} provides a practical demonstration of our method's advantage over other PR initializations in this regard.  The posterior for this injection is bimodal in the inclination angle, which the \texttt{simple-pe} analysis successfully captures. This information is then passed to the repartitioned prior, which ensures the subsequent NS analysis explores both modes and converges to the correct, unbiased posterior distribution. This case highlights how the physically-motivated model of \texttt{simple-pe} can provide a more reliable starting point for PR than a simple stochastic search, preserving the statistical integrity of the analysis for complex posterior geometries.

In summary, the injection study successfully validates that our method produces statistically consistent posteriors across the considered parameter range, and demonstrates the potential performance improvements.


\section{SNR Scaling Study}
\label{sec:snr_scaling_study}

The efficiency of PR is determined by the quality of the initial posterior approximation. Since both the precision of \texttt{simple-pe} and the concentration of the true posterior are expected to improve at higher SNR, we anticipate that the computational speedup of the \texttt{simple-pe}-PR algorithm will grow with signal strength. This was empirically supported by the results in Section~\ref{sec:injection_study}, but now we conduct a targeted study to investigate the performance of our method for signals with varying SNR. 

We focus on a single fiducial binary black hole merger. The parameters for this injection are listed in Table~\ref{tab:snr_study_injection_params}, and were chosen randomly from the prior specified in Table~\ref{tab:priors}. We then generated a series of signals, each 8 seconds in duration. For each signal, all source parameters were kept constant except for the luminosity distance, which was scaled to produce signals with target network SNRs of 10, 20, 30, 40, 50, 75, 100, and 150.

\begin{table}[t!]
    \centering
    \begin{tabular}{l c}
        \toprule
        \textbf{Parameter} & \textbf{Value} \\
        \midrule
        Chirp mass ($\mathcal{M}_c$)        & $22.5\, \textrm{M}_\odot$ \\
        Mass ratio ($q$)                    & $0.11$ \\
        Right ascension ($\alpha$)          & $1.11\, \textrm{rad}$ \\
        Declination ($\delta$)              & $1.33\, \textrm{rad}$ \\
        Inclination angle ($\theta_\textrm{JN}$) & $2.59\, \textrm{rad}$ \\
        Polarization angle ($\psi$)         & $1.01\, \textrm{rad}$ \\
        Phase at coalescence ($\phi$)       & $3.82\, \textrm{rad}$ \\
        Spin of lighter BH ($\chi_1$)       & $0.28$ \\
        Spin of heavier BH ($\chi_2$)       & $0.49$ \\
        Time of coalescence ($t_c$)         & $1367638517.66\, \textrm{s}$ \\
        \bottomrule
    \end{tabular}
        \caption{Fiducial detector frame parameters for the injection used in the SNR scaling study. The luminosity distance is not listed as it was varied to achieve the target SNRs.}
    \label{tab:snr_study_injection_params}
\end{table}

The injection and recovery of each signal followed the procedure detailed in Section~\ref{sec:injection_study}. We used the priors detailed in Table~\ref{tab:priors} in Appendix A, but adopted a wider luminosity distance prior, from 100 to 5000 Mpc, which was more suitable for the low-SNR injections in this study.


\subsection{Statistical Agreement}

\begin{figure}
    \centering
    \includegraphics{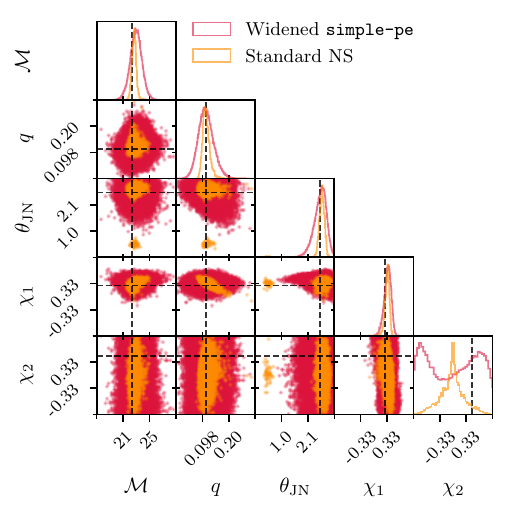}
    \caption{For the SNR 20 signal, \texttt{simple-pe} did not produce sufficient samples from the secondary mode of the true posterior in order for the flow to learn this feature. The widened flow is able to generate samples which lie in this secondary mode, but with very low probability. Since the sampler recognises that this region of the parameter space has a high repartitioned likelihood, it attempts to explore it, but inefficiently and unsuccessfully, leading a marginally slower analysis than standard NS and a slightly different final posterior. This issue can be mitigated by drawing more initial samples from \texttt{simple-pe} to make sure the secondary mode is sufficiently covered.}
    \label{fig:snr20}
\end{figure}

\begin{figure*}
    \centering
    \includegraphics{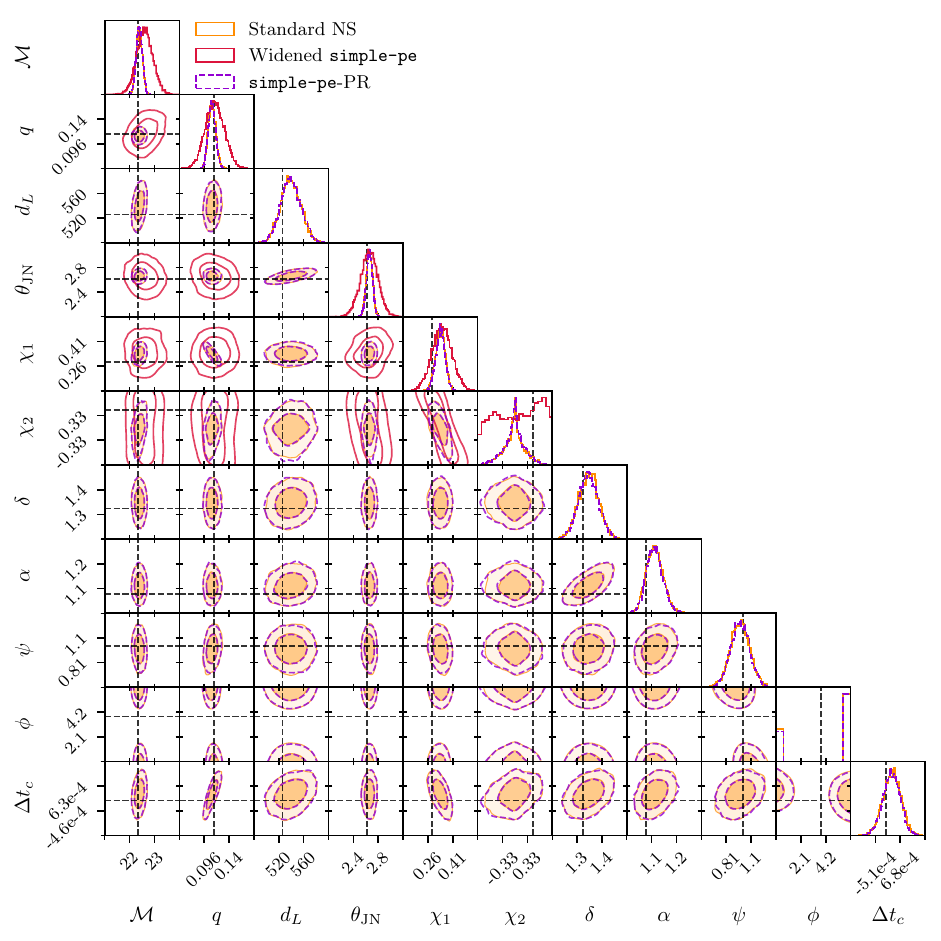}
    \caption{Validation of the \texttt{simple-pe}-PR method on the SNR 40 injection. The one- and two-dimensional marginal posterior distributions from our method (violet) are plotted alongside those from a standard NS analysis (orange), showing excellent agreement. The repartitioned prior, $\pi'$, used to accelerate the analysis is also shown for the five parameters to which it was applied (crimson).}
    \label{fig:fullposteriorsnr}
\end{figure*}

The posterior distributions produced by \texttt{simple-pe}-PR and standard NS showed excellent agreement for the signals with SNR $> 20$ (Figure~\ref{fig:fullposteriorsnr}). For the SNR 10 signal, our \texttt{simple-pe}-PR method produced a narrower posterior on the mass ratio, $q$, than standard NS, due to insufficient widening of the initial \texttt{simple-pe} distribution. For the SNR 20 signal, the standard NS analysis revealed a very small secondary mode in the inclination angle (containing 0.6\% of the samples) that was absent from our method's result. This feature was not present in the initial \texttt{simple-pe} estimate and was sufficiently distant from the primary mode that our widening procedure did not fully encompass it (see Figure~\ref{fig:snr20}). This exemplifies another limitation of the method: although \texttt{simple-pe} can handle multi-modal posteriors, as demonstrated in Section~\ref{sec:injection_study}, nevertheless it can still miss secondary features, particularly when their relative posterior mass is very small. This risk is most pronounced at low SNRs where posteriors can be complex, and is significantly mitigated for high SNR signals. Moreover, \texttt{simple-pe} produces samples drawn from a weighted distribution. The risk can be further mitigated, therefore, even at lower SNRs by drawing more samples to provide to the flow, which takes negligible additional time. Alternatively, the flow could also be trained directly on the weighted samples themselves. 

\subsection{Performance Scaling with SNR}

\begin{figure}
    \centering
    \includegraphics{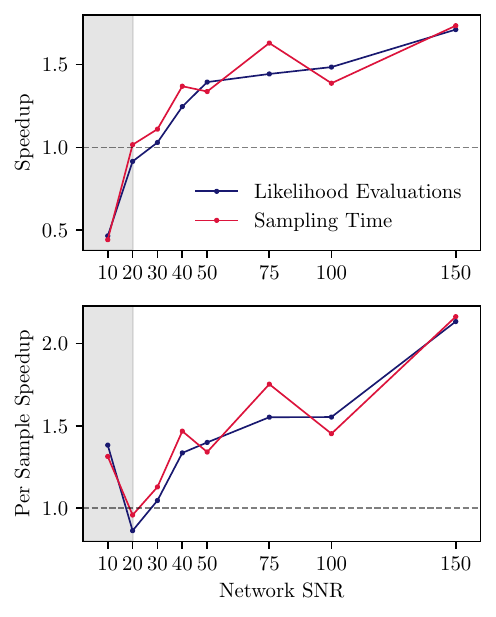}
    \caption{Performance of the \texttt{simple-pe}-PR method relative to a standard nested sampling analysis as a function of network SNR. The speedup factor is the ratio of a metric for the standard analysis to that of the PR analysis. \textit{Top:} Total speedup factors, showing the ratio of total likelihood evaluations (blue) and total sampling time (crimson). \textit{Bottom:} Speedup factors normalised by the number of effective posterior samples. The dashed line at 1.0 indicates no change in performance. The shaded and non-shaded regions separate the two different regimes used for the widening factors.}
    \label{fig:snr_scaling_plots}
\end{figure}

Figure~\ref{fig:snr_scaling_plots} shows the speedup factors as a function of optimal network SNR. Two metrics are considered when quantifying the performance improvement: the number of likelihood evaluations and the sampling time. Note that the speedup difference between these two metrics at a given SNR arises from the computational overhead of the repartitioned likelihood (Eq.~\ref{eq:repartitioned_likelihood}). This calculation requires evaluating not only the original likelihood, but also the log-probabilities of both the normalizing flow and the original prior. This additional cost would be mitigated if a vectorized sampler were used, which could evaluate the flow probabilities of multiple points at once, reducing the average single particle evaluation time~\citep{cabezas2024blackjax,flowMC,jaxns, yallup2025nested, Williams2021Nessai, Tiwari:2023mzf}. The top panel of Figure~\ref{fig:snr_scaling_plots} plots the overall speedup, whilst the bottom panel shows the per-sample speedup. 

In terms of overall speedup (top panel), our method is slower than the standard approach at SNR 10 and 20. As the SNR increases, however, the method's performance improves substantially. Above SNR 20, \texttt{simple-pe}-PR becomes more efficient than standard NS and by SNR 150, the number of likelihood evaluations and the total sampling time are reduced by a factor of 1.7, corresponding to a $41\%$ reduction. While the overall trend shows a clear monotonic increase in performance with SNR, minor deviations in the sampling time are consistent with temporal variations in CPU node performance on high performance computing systems.

Considering instead the per-sample speedup (bottom panel), the performance at low SNR is now more favourable; only the SNR 20 analysis is marginally slower. At SNR 150, the number of likelihood evaluations and total sampling time are now reduced by a factor of 2.1 (a $53\%$ reduction). 
Furthermore, the per-sample speedup at high SNR shows no sign of saturating, suggesting that further efficiency gains may be possible for even louder signals. Although SNR 150 signals are unlikely to be routinely observed until the advent of third-generation detectors or LISA~\citep{Punturo:2010zza,Hild:2010id,ET:2019dnz,Babak:2021mhe,LISA:2024hlh}, our method still provides an improvement on standard NS above SNR 20. We already observe speedups of above 1.5 at SNR $>50$, a regime where signals are starting to be observed in O4~\citep{LIGOScientific:2025rid, LIGOScientific:2025obp} and will become more frequent in O5.

At SNR 10, the poor performance of our method can be explained by the posterior distribution itself being broader at low SNR. At SNR 20, the missed secondary mode in the widened flow leads to sampling inefficiencies. The sampler attempts to explore the part of the parameter space containing the secondary mode, recognising it as a region of high likelihood for the repartitioned analysis (Eq.~\ref{eq:repartitioned_likelihood}), but since this lies deep in the tails of the widened flow, the analysis requires more likelihood evaluations (Figure~\ref{fig:snr20}).

\begin{figure}
    \centering
    \includegraphics{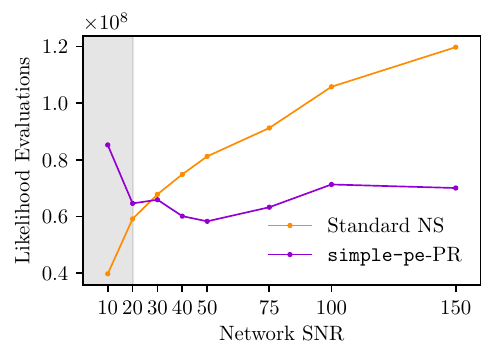}
    \caption{Total number of likelihood evaluations as a function of network SNR for a standard analysis (orange) and our \texttt{simple-pe}-PR analysis (violet). The computational cost of the standard analysis grows with SNR, while the cost of the PR analysis remains approximately constant for SNR $>$ 20.}
    \label{fig:likelihood_scaling}
\end{figure}

\begin{figure}
    \centering
    \includegraphics{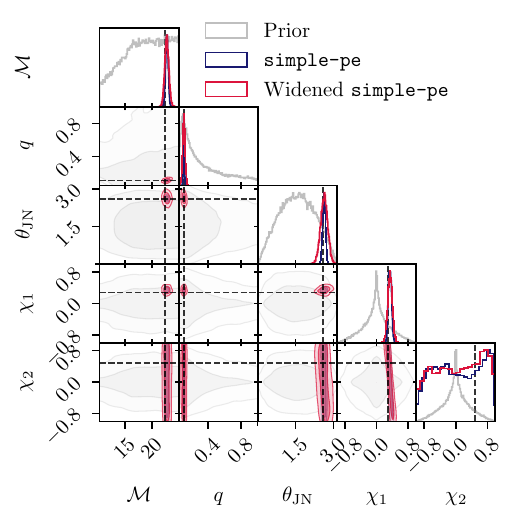}
    \caption{The prior, $\pi$, \texttt{simple-pe} distribution and widened distribution used as the repartitioned prior, $\pi'$, are shown for the SNR 40 injection. The repartitioned prior occupies a much smaller volume than the original prior, reducing the number of likelihood evaluations required for convergence. In all parameters but the spin of the heavier BH ($\chi_2$), the repartitioned prior is much more constrained than the original prior and effectively removes the initial `search phase' of NS.}
    \label{fig:priortosimplepe}
\end{figure}

The origin of our method's performance improvement beyond SNR 20 is illustrated in Figure~\ref{fig:likelihood_scaling}. As expected, the number of likelihood evaluations required by the standard NS analysis grows with SNR. This is because the posterior volume becomes increasingly small relative to the prior, increasing the number of iterations of the algorithm required for convergence. In contrast, the number of likelihood evaluations for the PR analysis shows no strong scaling with SNR for signals above SNR 20, remaining approximately constant within the stochastic variations expected from the normalizing flow training and the sampling process. Our method effectively removes the expensive initial `search phase' of the nested sampling algorithm (see Figure~\ref{fig:priortosimplepe}). 

This efficiency gain has significant practical implications. As well as reducing runtimes by hours to days, this reduction in cost lowers the barrier for performing analyses with more physically complete but expensive waveform models, such as those including orbital eccentricity~\citep{Ramos-Buades:2021adz,Gamboa:2024hli,Nagar:2024dzj,Planas:2025feq}. Although many of the signals in version 4.0 of the Gravitational-Wave Transient Catalog (GWTC-4.0) have SNRs $<$ 20, 6\% have SNRs $>$ 20~\citep{LIGOScientific:2025slb}, and this percentage is likely to increase as detector sensitivity improves~\cite{aLVK_prospects}. 

Overall, the performance of our method is primarily driven by the Kullback-Leibler (KL) divergence between the prior and the posterior, which can be thought of as the `amount of compression' required for the sampler to converge~\citep{skilling, supernest, Prathaban}. This depends most strongly on the SNR of the signal and the initial prior volume. For example, for a low-SNR signal this divergence is modest, and the standard NS analysis already converges relatively quickly. In this regime, applying a conservative widening factor to the already broad \texttt{simple-pe} approximation can increase the effective KL divergence, leading to a longer runtime (see Appendix~\ref{appendix:kldivs} for further discussion). Additional sampling inefficiencies can be introduced if the widened flow proposes samples outside the support of the original prior (see  Appendix~\ref{appendix:wideningflows} for further discussion). We do not expect the speedups to vary significantly with the binary's component masses independently; any observed correlation with mass is an indirect consequence of these two dominant factors. This conclusion relies on the assumption that the quality of the initial \texttt{simple-pe} approximation is reasonably consistent across the parameter space. While a detailed study of \texttt{simple-pe}'s performance is beyond the scope of this paper, our findings indicate that it is likely a subdominant effect regardless. 
For the injections in this section, the \texttt{simple-pe} approximation became more accurate at higher SNRs, yet we achieved significant speedup improvements beyond SNR 20 without changing the multiplicative widening factors. This demonstrates that the dominant driver of the acceleration is the overall reduction in the prior-to-posterior volume, rather than marginal improvements in the precision of the initial guess.

\section{Conclusions}
\label{sec:conclusions}

The increasing detection rate and SNR of gravitational-wave events from the global detector network are enabling an era of high-precision science, but the computational cost of the required parameter estimation presents a significant challenge~\citep{aLIGO, aVirgo, aLVK_prospects}. Standard nested sampling, while robust, is computationally expensive, particularly for the loudest signals. In this work, we have presented and validated a method to accelerate this process by combining the rapid, physically-motivated constraints from \texttt{simple-pe} with the nested sampling acceleration technique of posterior repartitioning.

An injection study, performed on a population drawn from standard analysis priors for 8 second signals, rigorously validated the statistical integrity of our method. The successful probability-probability test and the detailed analysis of challenging cases confirm that our method produces unbiased posteriors consistent with standard techniques, correctly prioritising statistical robustness over speed when necessary.

We have established a clear relationship between SNR and performance, showing that the computational speedup increases with SNR. For a fiducial signal at an SNR of 150, we achieve a per-sample speedup factor of approximately 2.1 in the number of likelihood evaluations. By comparison, a signal produced by the same parameters but a smaller luminosity distance, with SNR 30, achieves a per-sample speedup of 1.2, showing that the efficiency grows significantly with signal strength. Furthermore, the per-sample speedup does not show signs of saturating at SNR 150, indicating that further gains are possible for yet louder signals. This positions our method as a powerful algorithm for next-generation signals from observatories like the Einstein Telescope and LISA~\citep{Punturo:2010zza,Hild:2010id,ET:2019dnz,Babak:2021mhe,LISA:2024hlh}. 

Beyond the numerical speedup, our method provides a way to capitalise on the standard GW analysis workflow. Low-latency analyses, like \texttt{simple-pe}, are essential for rapid event alerts, but the valuable information they provide about the likely region of parameter space is typically set aside when the final, high-fidelity parameter estimation is performed. This separation is necessary to preserve the integrity of the Bayesian prior, which must not be informed by the data being analysed. Our method provides an innovative strategy for this purpose. As a purely computational technique, it uses the low-latency result to focus the sampler on the region of interest, while the repartitioning factor (Eq.~\ref{eq:repartitioned_likelihood}) mathematically ensures that the final posterior is identical to one produced from the original, unaltered prior. The method thus gains the efficiency of an informed analysis while rigorously preserving the statistical purity of the final inference.

The repartitioning algorithm presented here is not specific to \texttt{simple-pe}. It is possible to use our technique with machine learning algorithms, which also rapidly produce initial posterior estimates~\cite{Gabbard:2019rde,Green:2020dnx,Green:2020hst,Dax:2021tsq,Dax:2024mcn,Raymond:2024xzj}. However, care should be taken when using initialization methods which are not robust to multi-modal posteriors, such as a Fisher matrix approximation, as these could cut off secondary modes in the subsequent analysis and bias the inference. Our method is also complementary to likelihood acceleration techniques; \texttt{simple-pe}-PR reduces the number of required likelihood evaluations, and thus can be combined with optimised likelihoods that reduce the cost of each individual evaluation for even further speedups~\citep{Cornish:2021lje, Canizares:2014fya, Multibanding1, Multibanding2}. \texttt{simple-pe} could also be used to seed alternative sampling methods besides NS, such as Sequential Monte Carlo, to accelerate GW inference~\cite{SMC_GW, SMC_accelerated}. An important future direction for this work is in automating the widening procedure, in order to improve both the robustness and the performance gains. Moreover, our method could be extended to explore precessing systems.

In summary, we have developed a robust algorithm for accelerating gravitational-wave parameter estimation in the high-SNR regime. This method focuses on speeding up the core nested sampling algorithm itself, meaning it can be used in conjunction with other established acceleration techniques for further gains. This work represents a substantial step towards preparing for the data analysis challenges of the next generation of gravitational-wave observatories and will help to realise the full scientific potential of future detections.

\begin{acknowledgments}

We thank Sama Al-Shammari for helpful comments during the LIGO--Virgo--KAGRA internal review. We additionally thank Stephen Fairhurst, Laura Nuttall and Mukesh Singh for valuable discussions, and Stephen Fairhurst and Mukesh Singh for their continued development of {\texttt{simple-pe}} (alongside C.H).
M.P was supported by the research environment and infrastructure of the Handley Lab at the University of Cambridge. C.H thanks the UKRI Future Leaders Fellowship for support through the grant MR/T01881X/1, and the University of Portsmouth for support through the Dennis Sciama Fellowship. M.J.W acknowledges support from ST/X002225/1, ST/Y004876/1 and the University
of Portsmouth. For the purpose of open access, the author(s) has applied a Creative Commons Attribution (CC BY) licence to any Author Accepted Manuscript version arising.

This work was performed using the Cambridge Service for Data Driven Discovery (CSD3), part of which is operated by the University of Cambridge Research Computing on behalf of the STFC DiRAC HPC Facility (www.dirac.ac.uk). The DiRAC component of CSD3 was funded by BEIS capital funding via STFC capital grants ST/P002307/1 and ST/R002452/1 and STFC operations grant ST/R00689X/1. DiRAC is part of the National e-Infrastructure.

This work made use of \texttt{numpy}~\citep{harris2020array}, \texttt{scipy}~\citep{2020SciPy-NMeth}, \texttt{simple-pe}~\citep{SimplePe1}, {\texttt{pesummary}}~\citep{Hoy:2020vys}, \texttt{bilby}=2.3.0~\citep{bilby_paper, bilby_v2.4.0}, \texttt{anesthetic}~\citep{anesthetic} and \texttt{margarine}~\citep{margarine1, margarine2}. Plots were prepared with \texttt{matplotlib}~\citep{2007CSE.....9...90H} and \texttt{anesthetic}~\citep{anesthetic}.

\end{acknowledgments}

\clearpage
\appendix

\section{Priors}
\label{appendix:priors}

The prior distributions used for the main nested sampling analyses, detailed in Table~\ref{tab:priors}, are chosen to reflect the common priors used by the LVK~\citep{LIGOScientific:2025slb} and such that the gravitational-wave signal fits within the specified 8s data duration. This is commonly done in the literature (see e.g.~\cite{PhenomXPNR}). This includes priors that are uniform in the component masses, though the sampler itself explores the parameter space in terms of chirp mass and mass ratio. In contrast, the \texttt{simple-pe} analysis that initialises our method uses priors uniform in chirp mass and mass ratio directly. Similarly, the sampling priors on the spins are distributed uniformly over the sphere, whereas \texttt{simple-pe} uses a uniform prior on the effective aligned spin of the binary.

While this mismatch does not affect the statistical validity of the final posterior, our investigations show that it may have an impact on the computational efficiency of the method. The initial posterior approximation from \texttt{simple-pe} is conditioned on its own prior. Therefore, when the final analysis uses a different prior, the \texttt{simple-pe} guess can be more offset from the target posterior. This may necessitate more significant widening of the repartitioned prior and results in a larger effective KL divergence for the sampler to traverse, reducing the overall speedup.

Improved performance could be achieved in future work by using consistent priors throughout the method. This could be accomplished by adapting \texttt{simple-pe} to use component-mass priors or by implementing a prior re-weighting scheme for the initial samples before they are used to train the normalizing flow.

\begin{table}
    \centering
    \begin{tabular}{l c l}
        \toprule
        \textbf{Parameter} & \textbf{Shape} & \textbf{Range} \\
        \midrule
        Chirp mass ($\mathcal{M}_c$)        & Uniform (comp)     & $[10, 25]\, \textrm{M}_\odot$ \\
        Mass ratio ($q$)                    & Uniform (comp)        & $[0.05, 1.0]$ \\
        Luminosity distance ($d_L$)         & Uniform (source) & $[100, 1000]\, \textrm{Mpc}$ \\
        Right ascension ($\alpha$)          & Uniform        & $[0, 2\pi]\, \textrm{rad}$ \\
        Declination ($\delta$)              & Cosine         & $[-\pi/2, \pi/2]\, \textrm{rad}$ \\
        Inclination angle ($\theta_\textrm{JN}$)         & Sine           & $[0, \pi]\, \textrm{rad}$ \\
        Polarization angle ($\psi$)         & Uniform        & $[0, \pi]\, \textrm{rad}$ \\
        Phase at coalescence ($\phi$)       & Uniform        & $[0, 2\pi]\, \textrm{rad}$ \\
        Spin of lighter BH ($\chi_1$)  & Uniform (aligned)       & $[-0.99, 0.99]$ \\
        Spin of heavier BH ($\chi_2$) & Uniform  (aligned)      & $[-0.99, 0.99]$ \\
        Time of coalescence ($t_c$)         & Uniform        & \parbox[t]{2.5cm}{trigger time + \\ $[-0.1,+0.1]$ s}  \\
        \bottomrule
    \end{tabular}
     \caption{Prior distributions for the 100 injections used in the injection study. The choice of a source-frame uniform luminosity distance prior and uniform component (comp) mass priors are chosen to be consistent with standard LVK analyses. }
    \label{tab:priors}
\end{table}

\section{Empirical Calibration of the Widening Factors}
\label{appendix:wideningflows}

To ensure the repartitioned prior robustly encompasses the true posterior, the initial distribution from \texttt{simple-pe} must be widened. We detail here the empirical procedure used to calibrate the widening factors applied throughout this work.

The calibration was performed using the 100-injection set described in Section~\ref{sec:injection_study}. For each injection, we determined the minimum widening required for the repartitioned prior to be statistically consistent with the posterior from the standard NS analysis. The procedure was as follows. First, a normalizing flow was trained on the initial \texttt{simple-pe} samples. Then, the standard deviation of each diagonal element of the flow's base covariance matrix was systematically increased. This process was repeated until the 1st and 99th percentiles of the standard NS posterior were contained within the 1st and 99th percentiles of the distribution generated by the widened flow. This calibration was performed independently for each repartitioned parameter.

\begin{figure}
    \centering
    \includegraphics{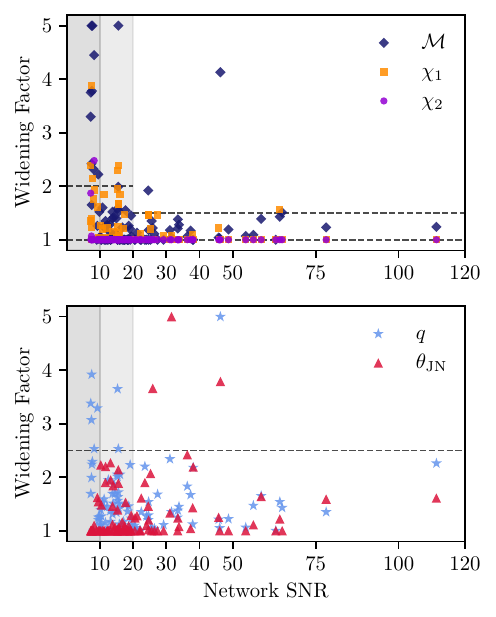}
    \caption{The minimum widening factor required for the repartitioned prior to encompass the 1st and 99th percentiles of the standard nested sampling posterior for each of the 100 injections. The widening factor represents the multiplicative value applied to the standard deviation of the normalizing flow's base distribution for a given parameter. These results were used to calibrate the two sets of fixed, conservative factors used in this work (dashed lines).}
    \label{fig:snr_widening}
\end{figure}

The minimum required widening factor for each parameter is shown as a function of network SNR in Figure~\ref{fig:snr_widening}. The results show a clear dependence on SNR for several parameters, particularly the chirp mass and spins, with a notable change in behavior around an SNR of 20. For many signals, especially those above SNR 10, little to no widening was required. Based on these results, we adopted the two sets of fixed, conservative widening factors detailed in Table~\ref{tab:widening_low_snr}.

These widening factors were chosen to be robust against the small, expected variations between the \texttt{simple-pe} approximation and the true posterior. We note that these factors are not tuned to accommodate rare outlier cases where the initial guess is significantly offset (as in the examples discussed in Section~\ref{sec:injection_study}). A more advanced implementation could employ an automated widening procedure, where the widening factors themselves are sampled during the analysis. This would be robust to outliers, applying large widening only when needed, while retaining higher efficiency for typical cases where minimal widening is sufficient. We leave the development of such a widening procedure for future work.

\section{KL Divergences and PR Speedups}
\label{appendix:kldivs}

The performance scaling of our method, presented in Section~\ref{sec:snr_scaling_study}, can be understood better through considering the Kullback-Leibler (KL) divergence. In the context of nested sampling, the KL divergence between the sampling prior and the posterior, $D_\textrm{KL}(\mathcal{P} || \pi)$, quantifies the information gain, or the `amount of compression', required for the sampler to converge from the prior to the posterior~\citep{skilling, supernest, Prathaban}. A larger KL divergence generally corresponds to a longer and more computationally expensive nested sampling run. By design, our method aims to reduce this divergence by replacing the broad original prior with a more compact repartitioned prior that is already a closer approximation of the posterior.

\begin{figure}
    \centering
    \includegraphics{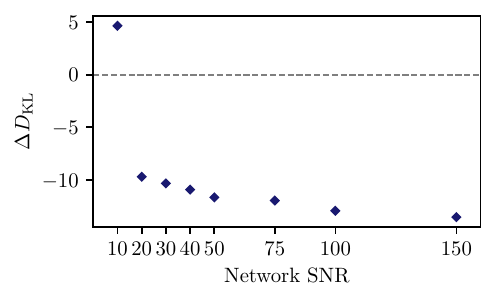}
    \caption{The change in the KL divergence between the sampling prior and posterior for our method relative to the standard NS analysis. A negative value indicates a reduction in the KL divergence for our method and an expected speedup. The plot shows that the method increases the required `compression' at SNR 10 but successfully reduces it for all higher SNRs, providing an explanation for the performance trend observed in Section~\ref{sec:snr_scaling_study}.}
    \label{fig:kl_div_vs_snr}
\end{figure}

Figure~\ref{fig:kl_div_vs_snr} plots the change in the KL divergence between our method and the standard NS analysis for the SNR scaling study. The results provide further insight into the observed performance.
At SNR 10, the substantial widening applied to the already broad \texttt{simple-pe} guess results in a repartitioned prior that is a poorer representation of the posterior than the original prior. This causes the KL divergence to increase, leading to the performance reduction reported in Section~\ref{sec:snr_scaling_study}.
For SNR $\geq$ 20, our method successfully reduces the KL divergence. However, at SNR 20 specifically, this gain is offset by the sampling inefficiency introduced by the missed secondary posterior mode, which explains the slight performance reduction in this case.
For SNR $\geq$ 30, the performance gains are driven directly by the reduction in KL divergence. The growing difference in KL divergence between our method and the standard analysis as SNR increases is the underlying reason for the trend of increasing speedups.

\section{Jensen-Shannon Divergences}
\label{appendix:jsd}

To quantitatively assess the statistical consistency between the posteriors from standard NS and our \texttt{simple-pe}-PR method, we compute the Jensen-Shannon Divergence (JSD) for the one-dimensional marginal distributions. The JSD is a measure of the distance between two probability distributions, taking values between 0 and $\ln2$ (1) nats (bits); 0 represents exactly identical distributions and a larger JSD indicates greater disagreement.  

To estimate the mean and uncertainty of the JSD for each comparison, we generate 100 posterior realisations per analysis. This is achieved by resampling the nested sampling weights associated with the full set of dead points, using the functionality in \texttt{anesthetic}~\citep{anesthetic}. From each set of newly weighted points, a final realisation of 5000 posterior samples is produced via rejection sampling. The JSD is then calculated between kernel density estimates of these marginal distributions, constructed using \texttt{PESummary}~\citep{pesummary}. JSD values below $0.05$ bits, or $35$ millinats, indicate that the distributions are statistically consistent~\citep{GWTC1}.

The JSD values for the SNR scaling study are presented in Table~\ref{tab:JSD_snr}. For all signals with SNR $>$ 10, the JSDs are below this threshold, confirming excellent agreement between the methods. As expected from the discussion in Section~\ref{sec:snr_scaling_study}, the SNR 10 signal exhibits a JSD value above the threshold, for the mass ratio parameter, providing a quantitative measure of the observed disagreement.

\begin{figure}
    \centering
    \includegraphics{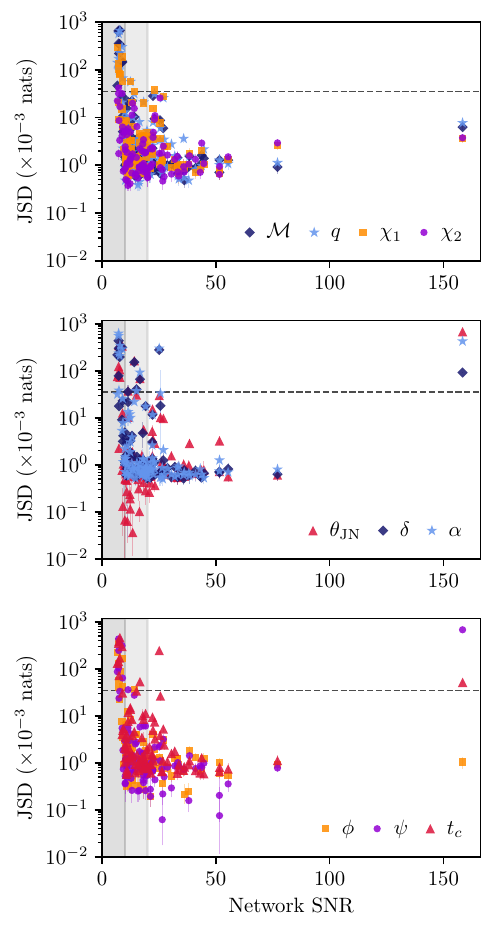}
    \caption{The Jensen-Shannon divergence (JSD) between the one-dimensional marginal posteriors from our method and the standard NS analysis for each of the 100 injections. Each point represents the JSD for a single parameter and for a single event. The horizontal dashed line at 35 millinats represents the threshold for statistical consistency~\citep{GWTC1}. The plot shows a clear trend of improved agreement at higher SNR.}
    \label{fig:jsd_vs_snr}
\end{figure}

The JSD values for the full injection study are presented in Figure~\ref{fig:jsd_vs_snr}. The results show a clear trend with SNR and provide quantitative support for the conclusions drawn in the main text. Below SNR 10, several injections exhibit JSDs above the 35 millinat threshold for consistency. This is expected, as our empirical widening factors were not calibrated for this low-SNR regime and can be insufficient to fully encompass the true posterior. In the 10 $\le$ SNR $<$ 20 regime, the agreement improves significantly, though a few analyses still produce posteriors that are statistically distinguishable from the standard result. For signals with SNR $>$ 20, only two injections fail the JSD test. These correspond to the specific outlier cases identified in Section~\ref{sec:injection_study}, where a significant offset in the initial \texttt{simple-pe} guess was not fully corrected by the widening procedure. An automated widening scheme would likely resolve these remaining discrepancies. These results provide quantitative evidence for our recommendation to apply the method to signals with SNR $>$ 20, the regime where it is both computationally efficient and statistically robust.

\begin{table*}[t!] 
    \centering
    \includegraphics{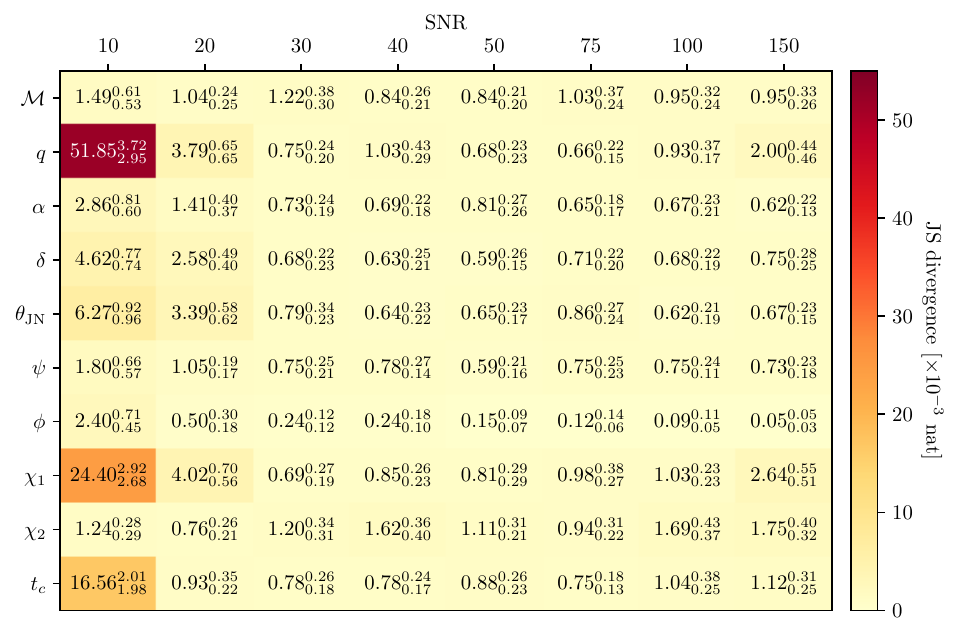}
    \caption{Jensen-Shannon divergences for marginal posterior distributions between standard NS and our method, from the SNR scaling study. The quoted values are in units of $1 \times 10^{-3}$ nats. 100 different realisations of 5000 posterior samples were generated for each analysis by resampling the nested sampling weights and performing rejecting sampling again. The mean and $1\sigma$ quantiles are shown, where for injections above SNR 10 the JSD values are within the threshold of $0.05$ bits, or $35$ millinats~\cite{GWTC1}. There are significant disagreements for the SNR 10 signal for the mass ratio parameter, but we do not recommend the application of our method in this SNR regime. }
    \label{tab:JSD_snr}
\end{table*}

\section{Effective Sample Size}
\label{appendix:ess}

The effective sample size from a nested sampling analysis can be obtained using the set of nested sampling weights, $w_i$, for each dead point, $i$. This can be calculated as~\citep{Kish1995}:

\begin{equation}
    n_\textrm{eff} = \frac{(\sum_i w_i)^2}{\sum_i w_i^2}.
\end{equation}

\begin{figure}
    \centering
    \includegraphics{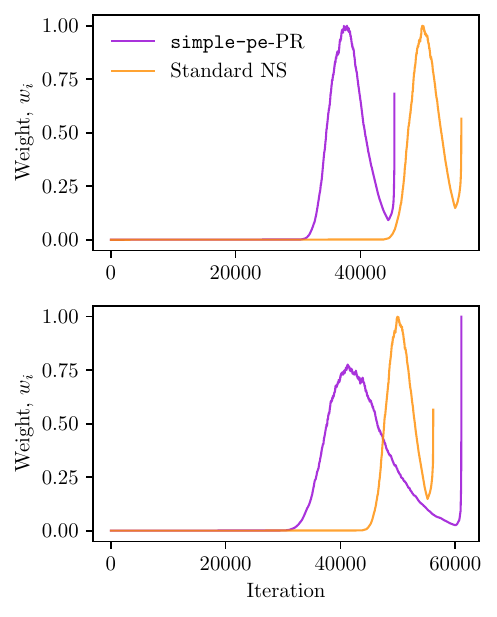}
    \caption{Comparison of the posterior weight distribution of the dead points for a standard NS analysis (orange) and our \texttt{simple-pe}-PR method (violet). The x-axis represents the nested sampling iteration, and the y-axis is the posterior weight. The top panel shows a typical case where the PR analysis enters the high-likelihood region (where weights become non-negligible) earlier but the phase of significant weight extends over slightly more iterations, resulting in a larger effective sample size. The bottom panel shows a more extreme case where the compression through the posterior is much slower for the PR analysis, leading to a substantially larger effective sample size at the cost of a longer runtime. Both cases demonstrate that a direct comparison of runtimes at a fixed number of live points is misleading, and the per-sample speedup is a more informative metric.}
    \label{fig:pr_weights}
\end{figure}

Posterior repartitioned NS typically spends a longer time exploring the bulk of the posterior than standard NS for a given number of live points (see Figure~\ref{fig:pr_weights}). This can be explained by PR having a greater density of points in this region, due to the smaller effective prior volume. Thus, the rate of prior volume compression occurs more slowly in this region, generating a larger number of dead points with non-negligible weights, and therefore resulting in a larger effective sample size. As a result, our method should be performed with fewer live points than a standard NS analysis in order to obtain a given final sample size.

\section{P-P Plots}
\label{appendix:ppplot}

\begin{figure*}
    \centering
    \includegraphics{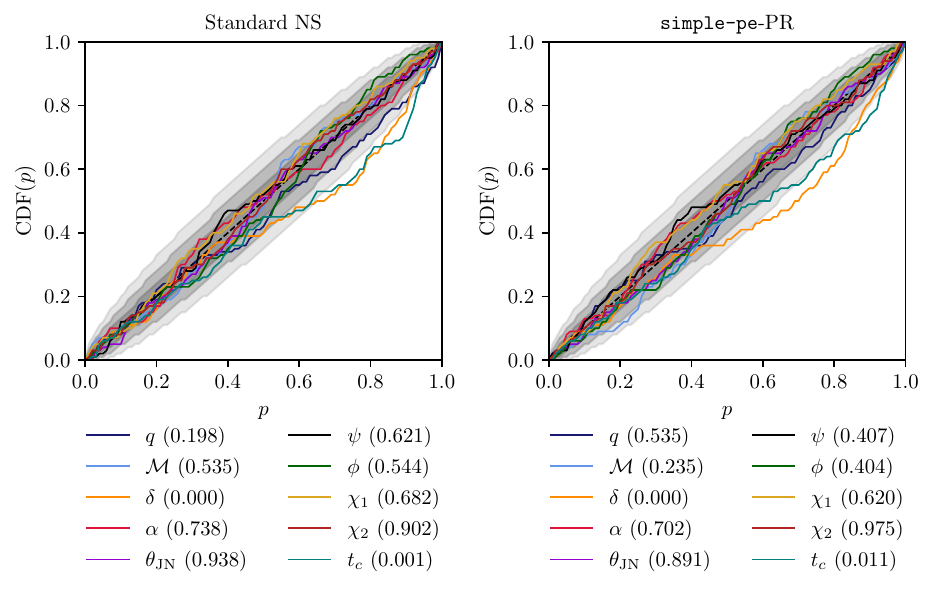}
    \caption{Probability-probability (P-P) plot for the 100 injections in our validation study, analysed using both standard NS and \texttt{simple-pe}-PR. Each coloured line represents the cumulative distribution function (CDF) of the posterior quantiles of the true injected values for a specific parameter. The diagonal dashed line represents a perfect uniform distribution, while the shaded gray regions indicate the $1\sigma$, $2\sigma$ and $3\sigma$
 confidence intervals. The p-values from a Kolmogorov-Smirnov test for each parameter are shown in the legend. We observe biases in the declination and time of coalescence at the geocentre for both methods, which we attribute to statistical fluctuations in drawing our injections, since this bias was not observed in the calibration set. Importantly, our algorithm produces percentiles consistent with those from standard NS, demonstrating agreement between the two methods on a population level.}
    \label{fig:pp_plot}
\end{figure*}

\begin{figure}
    \centering
    \includegraphics{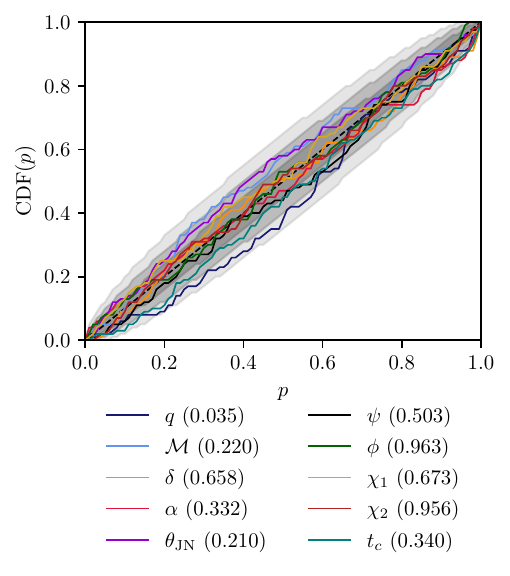}
    \caption{P-P plot obtained from the standard nested sampling analyses performed on our calibration set from Section~\ref{sec:calib} using the \texttt{dynesty} sampler.}
    \label{fig:normal_ppplot}
\end{figure}

\begin{figure}
    \centering
    \includegraphics{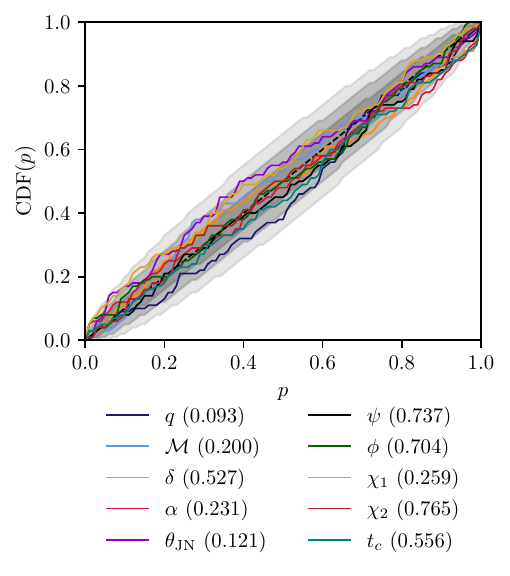}
    \caption{P-P plot obtained from the \texttt{simple-pe}-PR analyses performed on our calibration set from Section~\ref{sec:calib} using the \texttt{dynesty} sampler, modified to implement PR.}
    \label{fig:pr_ppplot}
\end{figure}

\begin{figure*}
    \centering
    \includegraphics{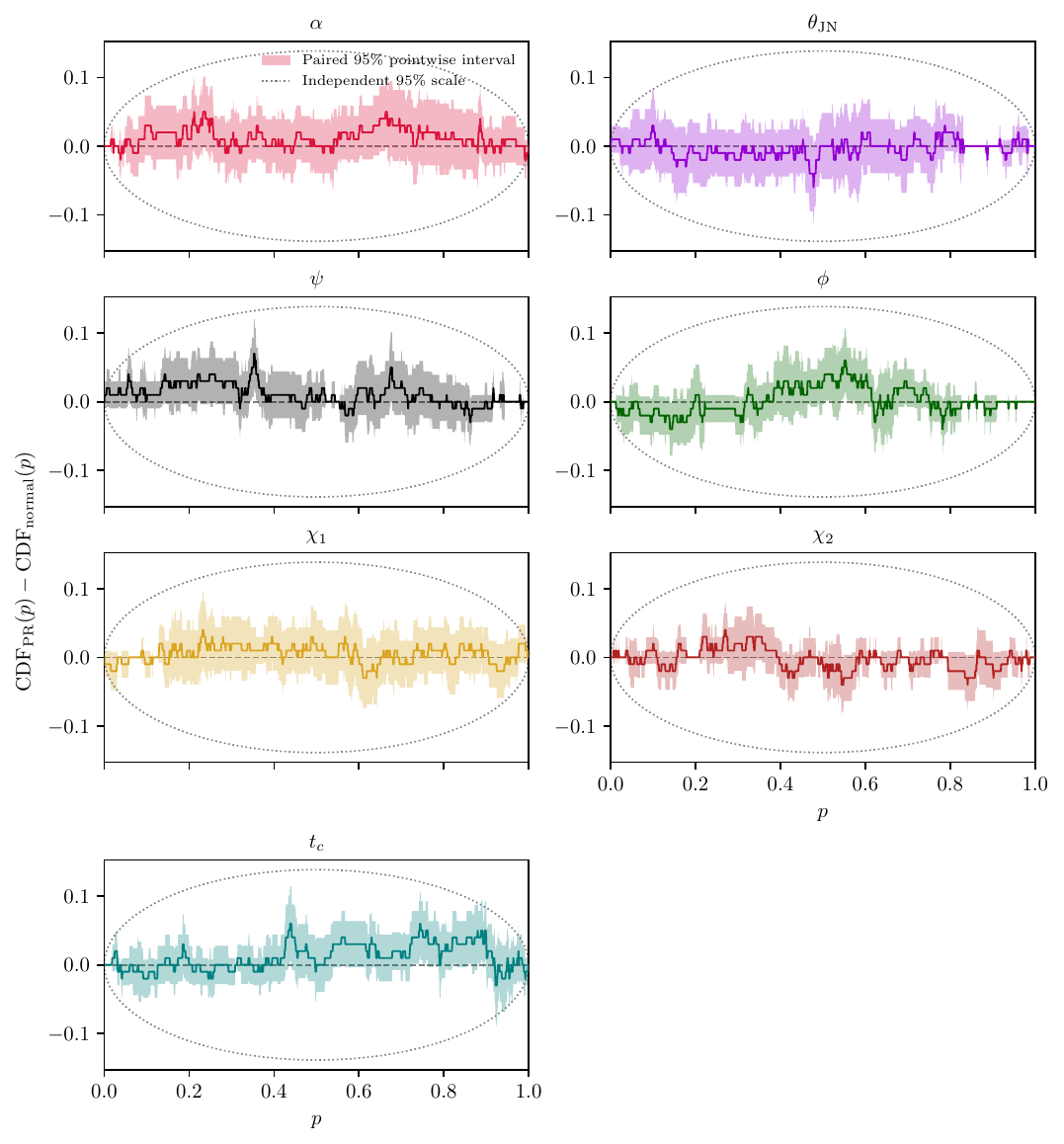}
    \caption{The differences between the P-P plots obtained from \texttt{simple-pe}-PR and normal NS are shown for the remaining parameters (solid). Perfect agreement corresponds to a horizontal line at zero, but there is statistical uncertainty; within the error bars (shaded), both methods agree at the population level, including the time of coalescence at the geocentre parameter, which exhibited small biases in the P-P plots from both methods.}
    \label{fig:pp_diff_remaining}
\end{figure*}

The P-P plots obtained from standard NS and \texttt{simple-pe}-PR on both the injection study set and the calibration set used to determine the necessary widening factors are included here for reference (Figures~\ref{fig:pp_plot},~\ref{fig:normal_ppplot} and~\ref{fig:pr_ppplot}). We note that in the case of the calibration set, neither P-P plot shows the same biases in declination and time of coalescence at the geocentre as seen for the injection set. This supports our hypothesis that the biases seen for the injection set can be attributed to statistical fluctuations in drawing the injections. In both cases, the P-P plot from our method is consistent with the P-P plot from standard NS, indicating agreement between the two methods. Finally, the difference plots for the remaining parameters not shown in the main text are also included here (Figure~\ref{fig:pp_diff_remaining}).


\bibliography{apssamp}

\end{document}